\documentclass[11pt]{article}

\usepackage[english]{babel}

\usepackage[a4paper,top=2cm,bottom=2cm,left=2.2cm,right=2.2cm,marginparwidth=1.75cm]{geometry}

\usepackage[utf8]{inputenc}
\usepackage[T1]{fontenc}

\usepackage{setspace}
\usepackage{graphicx}
\usepackage{amsmath,amssymb}

\usepackage[numbers]{natbib}
\usepackage{titlesec}
\usepackage{caption}
\usepackage{multicol}
\usepackage{ragged2e}
\usepackage[dvipsnames]{xcolor}
\usepackage{tcolorbox}
\usepackage{authblk}

\usepackage[colorlinks=true,allcolors=blue]{hyperref}

\makeatletter

\renewcommand\AB@affilsepx{, }

\def\myinlineaffils{%
  {\small
   \renewcommand{\and}{, }
   \AB@affillist}
}
\renewcommand{\maketitle}{%
  \begin{center}
    {\LARGE\@title\par}
    \vskip 1em
    {\large
      \setlength{\parskip}{0.3em}
      \setlength{\parindent}{0pt}
      \@author
      \par}%
    \vskip 1em
    \vskip 1em
  \end{center}
}
\makeatother

\title{Why the Northern Hemisphere Needs a 30–40\,m Telescope and the Science at Stake: Shaping Galaxies and Their Stars with Stellar Population Gradients, IMF Variations and Environmental Drivers in Cluster Early-Type Galaxies}

\author[1,2]{Vazdekis, A.}
\author[3,4]{S\'anchez-Bl\'azquez P.}
\author[1,2,5]{Ferr\'e-Mateu, A.}
\author[1,2]{Mart{\'{\i}}n-Navarro, I.}
\author[1,2,5]{Beasley, M.A.}
\author[1,2]{Aguerri, J.~A.~L.}
\author[3,4]{Camps-Fariña, A.}
\author[2,1]{de Lorenzo-C\'aceres, A.}
\author[6]{Eftekhari, E.}
\author[1,2]{Falc\'on-Barroso, J.}
\author[1,2,7]{Ferreras, I.}
\author[8]{La Barbera, F.}
\author[9]{Garc{\'{\i}}a-Benito, R.}
\author[9]{Gonz\'alez Delgado, R.~M.}
\author[10]{Longhetti, M.}
\author[11]{Maraston, C.}
\author[12,13]{P\'erez, I.}
\author[1,2]{Pinna, F.}
\author[14]{Quilis, V.}
\author[15]{Peletier, R.~F.}
\author[1,2,16]{S\'anchez-S\'anchez, S.~F}
\author[17]{Sansom, A.}
\author[18]{Scholz-D{\'{\i}}az, L.}
\author[19,20]{Spiniello, C.}
\author[21]{Thomas, D.}
\author[1,2]{Villaver, E.}
\affil[1]{Instituto de Astrof{\'{\i}}sica de Canarias, Calle V{\'{\i}}a L\'actea, s/n, E-38205 La Laguna, Tenerife, Spain}
\affil[2]{Departamento de Astrof{\'{\i}}sica, Universidad de La Laguna, E-38206 La Laguna, Tenerife, Spain}
\affil[3]{Departamento de F{\'{\i}}sica de la Tierra y Astrof{\'{\i}}sica, Universidad Complutense de Madrid, E-28040, Madrid, Spain}
\affil[4]{Institute of Particle Physics and Cosmos Science, IPARCOS, Plza Ciencias 1, E-28040, Madrid, Spain}
\affil[5]{Centre for Astrophysics and Supercomputing, Swinburne University, Australia}
\affil[6]{Leiden Observatory, Leiden University, P.O. Box 9513, 2300 RA Leiden, The Netherlands}
\affil[7]{Department of Physics and Astronomy, University College London, London WC1E 6BT, UK}
\affil[8]{INAF, Osservatorio Astronomico di Capodimonte, Salita Moiariello 16, 80131, Napoli}
\affil[9]{Instituto de Astrof{\'{\i}}sica de Andaluc{\'{\i}}a - CSIC, Glorieta de la Astronom{\'{\i}}a s/n, E-18008 Granada, Spain}
\affil[10]{INAF-Osservatorio Astronomico di Brera, via Brera 28, I-20121 Milano, Italy}
\affil[11]{Institute of Cosmology, University of Portsmouth, Burnaby Road, Portsmouth PO1 3FX, UK}
\affil[12]{Departamento de Física Te\'orica y del Cosmos, Universidad de Granada, E-18071, Granada, Spain}
\affil[13]{Instituto Carlos I de F{\'{\i}}sica Te\'orica y Computacional, Facultad de Ciencias, E-18071 Granada, Spain}
\affil[14]{Departament d'Astronomia i Astrof{\'{\i}}sica, Universitat de Val\`encia, E-46100 Burjassot, Val\`encia, Spain}
\affil[15]{Kapteyn Astronomical Institute, University of Groningen, The Netherlands}
\affil[16]{Instituto de Astronom{\'{\i}}a, Universidad Nacional Autonoma de Mexico, Mexico}
\affil[17]{Jeremiah Horrocks Institute, University of Lancashire, Preston, Lancashire, PR1 2HE, UK}
\affil[18]{INAF - Osservatorio Astrofisico di Arcetri
Largo Enrico Fermi, 5
50125 Firenze FI, Italy}
\affil[19]{Sub-Dep. of Astrophysics, Dep. of Physics, University of Oxford, Denys Wilkinson Building, Keble Road, Oxford OX1 3RH, United Kingdom}
\affil[20]{European Southern Observatory,  Karl-Schwarzschild-Stra\ss{}e 2, 85748, Garching, Germany}
\affil[21]{School of Physics and Astronomy, University of Leeds, Leeds, LS2 9JT, UK}
\begin{document}
\maketitle
\newpage
\begin{tcolorbox}[colback=RoyalBlue!5!white,colframe=black!75!black, width=\textwidth]
\justifying
\noindent {\em This white paper highlights how stellar population gradients, chemical abundance patterns, stellar initial mass function (IMF) variations, and structural signatures in early-type galaxies (ETGs), measured at faint and large galactocentric radii, out to $\sim4R_e$, provide powerful diagnostics of their formation and evolutionary histories. These observables encode the combined effects of early dissipative star formation, subsequent accretion and mergers, and internal feedback processes. Achieving such measurements requires high-signal-to-noise, spatially resolved U-band--optical--near-IR spectroscopy at large radii, with enough spatial resolution to study the variation of these properties on $\sim$kpc scales. These capabilities can only be delivered by a 30\,m-class telescope.
Disentangling these internal processes from environmental influences further demands observations of galaxies across clusters spanning a wide range of evolutionary stages and local environments. 
The nearby Virgo, Perseus, and Coma clusters, without any comparable nearby counterparts in the Southern Hemisphere, offer ideal laboratories for this work. Such observations will place stringent constraints on the formation mechanism of ETGs, connecting local cluster ETGs to their high-redshift progenitors. This white paper outlines several key science cases enabled by such a facility: (1) mapping stellar population gradients across environments; (2) tracing IMF variations as a function of evolutionary stage and environment; (3) reconstructing the three-dimensional structure of galaxies through deep integral-field spectroscopy and imaging; and (4) identifying and studying compact and relic systems as progenitors of present-day ETGs.}
\end{tcolorbox}


\section{Context and motivation}

Spatially resolved stellar population gradients in early-type galaxies (ETGs) provide powerful constraints on their formation and structural evolution, especially when studied across diverse environments. In nearby ellipticals, age, metallicity, and abundance-ratio gradients correlate with central velocity dispersion \citep[e.g.,][]{SanchezBlazquez07, MartinNavarro18}, reflecting the balance between early dissipative star formation and later internal growth. 
Although ETGs are dominated by old ($>10$~Gyr) stellar populations, with ages corresponding to those of massive ($\log_{10}(M_\star/M_\odot)>10$) galaxies at $z>3$ \citep[e.g.,][]{Carnall23, Glazebrook24}, their radial variations trace the imprint of feedback, gas accretion, and merger-driven assembly. Radial IMF variations, with M-dwarf–enhanced centers in the most massive systems \citep[e.g.,][]{MartinNavarro15, LaBarbera19}, further influence their inferred mass and enrichment histories.
Crucially, these processes depend sensitively on the surrounding environment: dense clusters such as Virgo, Perseus, and Coma accelerate quenching, enhance stripping, and modulate accretion and merger rates \citep[e.g.,][]{BoselliGavazzi06}. Small contributions from younger stars \citep[e.g.,][]{Trager00, Vazdekis16} complicate gradient interpretation, underscoring the need for high-precision measurements in well-characterized cluster settings.

The observed age gradients are mildly positive in the most massive ETGs, consistent with relatively more extended star formation in their centres, but become slightly negative at lower masses ($\log_{10}(M_\star/M_\odot) \sim 10$) \citep[e.g.,][]{PipinoMatteucci04, Pipino06}.
Metallicity gradients become steeper with increasing galaxy mass, as traced by central velocity dispersion \citep[e.g.,][]{KawataGibson03, ChiosiCarraro02}, suggesting that remnants of early in-situ formation survive in galaxy cores. The galaxy outskirts (including regions where shells, tidal tails, and other low-surface-brightness features reside) preserve the clearest signatures of environmental processing. Comparisons with compact massive relic galaxies, systems that represent the first phase of two-phase formation and have experienced little to no subsequent accretion \citep[e.g.,][]{FerreMateu17, MartinNavarro18}, show significantly flatter metallicity and [Mg/Fe] profiles in typical ETGs. This is consistent with environmentally driven minor mergers depositing chemically evolved, low-[Mg/Fe] material at large radii \citep{Cook16}, reinforcing a two-phase formation scenario shaped by internal processes and environment.

However, key questions remain. The partial decoupling of age and [Mg/Fe] gradients, the velocity-dispersion–dependent scatter in [Mg/Fe], all point to additional mechanisms, including IMF variations, affecting ETG chemical evolution \citep[e.g.,][]{
LaBarbera13, MartinNavarro16}. 
While [Mg/Fe] robustly traces enrichment timescales \citep[e.g.,][]{Vazdekis96, Feltzing01, Bensby14}, much less is known about other diagnostic element ratios, including CO-strong features in the NIR \citep[e.g.,][]{Silva08, Eftekhari22}. Importantly, [Mg/Fe] has now been measured at high redshift \citep[e.g.,][]{Lonoce15,Beverage21}, providing the opportunity to test whether observed gradients arise primarily from late-time accretion or are established mostly in-situ. Even small fractions of young stars leave strong UV signatures \citep[e.g.,][]{Kaviraj07, Yi05}, making combined multi-wavelength spectroscopy essential for reconstructing both chemical enrichment and star-formation histories.

Recent results from MaNGA \citep{Neumann21} indicate that stellar mass surface density, or even
local stellar velocity dispersion \citep{Ferreras25}, rather than radial position, predominantly drives the star formation histories within galaxies, with wide-reaching implications for accretion and galactic wind scenarios.  Hydrodynamical simulations, however, do not reproduce this pattern \citep{Nanni24}, highlighting the need to explore these trends across a broad range of environments. Because environmental effects regulate gas supply, merger activity, and stripping efficiency, disentangling in-situ and ex-situ growth requires spatially resolved measurements across diverse cluster conditions \citep[e.g.,][]{
Oser12, 
RodriguezGomez16}. To date, existing measurements of stellar population gradients do not reach these low-binding-energy outer regions, typically probing only the inner effective radius with few exceptions \citep[e.g.,][]{Greene15}. As a result, current observational constraints are insufficient to discriminate between competing formation pathways, such as the two-phase scenario, which predicts pronounced changes at large radii, and cannot robustly assess the impact of the environment, imprinted most strongly in the extreme outskirts. Reaching such faint surface brightnesses requires a 30\,m class telescope.


\section{Mapping Stellar Population Gradients and Galaxy Build-Up Across Cluster Environments and Evolutionary Stages}

Studying these processes demands observations extending beyond the effective radius in large galaxy samples, spanning rich clusters such as Virgo (dynamically young and nearby), Perseus (massive and star-forming rich), and Coma (compact, evolved, and dense). Critically, all three benchmark clusters lie in the Northern Hemisphere, with no equivalent ensemble of such nearby, massive, and morphologically diverse systems available in the South, having been extensively studied using 2--10\,m class telescopes. These clusters probe a wide spectrum of evolutionary stages and internal environments, from dense cores to infall regions and diffuse outskirts, allowing the response of stellar population gradients in ETGs to both global cluster properties and local environmental mechanisms to be quantified. In this context, simulations predict that local clusters should host the progenitors of ETGs that have largely avoided subsequent accretion, so-called relic galaxies \citep[e.g.,][]{QuilisTrujillo13,Stringer15}, although only a few have been identified, primarily in the Perseus Cluster \citep{FerreMateu17}. The absence of spatially resolved stellar population and kinematic studies for these rare systems highlights the need for systematic cluster surveys.

Tracing the high-$z$ build-up epoch of present day ETGs further requires sensitivity to faint spectral features across the U-band–NIR, which encode enrichment timescales and may vary systematically with environment. 
Discriminating between different models of galaxy formation requires reaching galactocentric distances  out to $\sim 4\,R_{\rm e}$ \citep{Cook16}, which is the regime where accreted stellar envelopes dominate and the predictions of galaxy formation models diverge \citep[e.g.,][]{Naab09, 
Hilz12}. Recovering the radial variation of metallicity, [Mg/Fe], and other key species, such as [Na/Fe] and [C/Fe] (see \cite{Johansson12}), which together trace the radial chemical enrichment history, along with IMF-sensitive features, is key to disentangling accretion from in-situ enrichment.
At $\sim 4\,R_{\rm e}$, surface brightnesses in massive cluster ETGs fall above $\mu_r \sim 25~{\rm mag\,arcsec^{-2}}$ (even fainter in the NIR), beyond the reach of current 8--10\,m facilities.

Northern access to Virgo, Perseus, and Coma ensures coverage of the most diverse and well-studied nearby cluster environments, crucial for isolating environmental effects on stellar population gradients. Combined with advanced integral-field spectroscopy, a 30\,m facility will provide the depth, wavelength coverage, and S/N to map gradients continuously from galaxy cores to $\sim 4\,R_{\rm e}$. Synergies with NIR instruments (e.g., EMIR, MOSFIRE) and space facilities (e.g., JWST) will extend the connection to high-redshift progenitors. By linking local cluster ETGs to their compact early-universe ancestors and mapping gradients with unprecedented fidelity, a next-generation 30\,m telescope will deliver decisive tests of galaxy formation models and an unparalleled view of how massive galaxies built their stars and metals over cosmic time.


\subsection{Galaxy Stellar Population and kinematic Gradients}
A multi-wavelength spectroscopic census of galaxy gradients across clusters at different evolutionary stages is essential for distinguishing competing growth scenarios. Gradients in age, metallicity, and elemental abundances encode the interplay between early dissipative processes, internal evolution, and environmental effects. High-S/N (S/N $\gtrsim 40$) spectroscopy at large radii ($R \sim 2$--$4\,R_{\rm e}$) across the U-band--optical--NIR will enable measurements of age- and metallicity-sensitive features, as well as key elemental abundance ratios. NIR stellar population diagnostics such as CO-bands or Na absorption in the K-band are highly sensitive to environment and intermediate-age populations \citep[e.g.,][]{Rock16, Eftekhari22}, while U-band data tightly constrain hot old or young components \citep[e.g.,][]{SalvadorRusinol21}. This combined coverage will refine stellar population models and reveal distinct evolutionary pathways.


\subsection{Initial Mass Function Within Galaxies}
The stellar IMF governs mass budgets and the chemical evolution of the Universe, gas thermodynamics, star and dust formation, and the build-up of stellar populations over cosmic time. Observations suggest increasingly bottom-heavy IMFs with increasing galaxy mass, and spatially resolved data reveal strong central IMF variations within ETGs \citep[e.g.,][]{MartinNavarro15}. Testing whether the IMF correlates with environment requires deep spectroscopy (S/N\,$\sim50-100$) near $\sim 1\,\mu{\rm m}$, where IMF-sensitive indices reside, and must reach the low-surface-brightness outskirts at $\sim4\,R_{\rm e}$; such observations are only feasible with 30\,m class telescopes. The Virgo, Perseus, and Coma clusters provide ideal laboratories to probe IMF variations across diverse environments, both within and among clusters, and to link IMF gradients with local star-formation conditions.


\subsection{Extreme Laboratories of Cluster-Driven Evolution}
Rare massive compact relic galaxies, local analogues of high-redshift red nuggets \citep[e.g.,][]{Buitrago08}, preserve ancient stellar populations, making them key probes of early formation. Their survival and evolution depend strongly on environment: dense cores may shield or erode them through tidal interactions. Conversely, lower-mass compact systems trace dynamical stripping and cluster tides. Deep spatially resolved spectroscopy across the full mass range, combined with gradient and IMF measurements, will elucidate the interplay between in-situ and environmental processes, revealing the diversity of evolutionary pathways in clusters and the role of environment in shaping galaxy structures.


\subsection{3D Views of Cluster Galaxies: Towards High-Precision Cosmology}
Integral-field spectroscopy and deep imaging enable reconstruction of the 3D structure of cluster galaxies by linking stellar populations, kinematics and morphology. Access to faint outskirts (including shells, tails and streams) provides evidence of past interactions, mergers and stripping. Combining structural diagnostics with stellar population and IMF gradients, reaching the semi-resolved regime, will provide a holistic view of galaxy build-up across environments, while mapping mass profiles and orbital anisotropies will inform dark matter models and allow high-precision cosmology.

\begin{multicols}{2}
\renewcommand{\bibfont}{\footnotesize}  
\bibliographystyle{unsrtnat}   
\bibliography{WP_GRADIENTS}    

@article{Bensby14,
    author  = "Bensby T. and Feltzing S. and Oey M.~S.",
    title   = "",
    year    = "2014",
    journal = "A\&A",
    volume  = "562",
    number  = "",
    pages   = "A71"
}

@article{Beverage21,
    author  = "Beverage A. and Kriek M. and Conroy C. and Bezanson R. and Franx M. and van der Wel A.",
    title   = "",
    year    = "2021",
    journal = "ApJ",
    volume  = "917",
    number  = "",
    pages   = "L1"
}

@article{BoselliGavazzi06,
    author  = "Boselli, A. and Gavazzi, G.",
    title   = "",
    year    = "2014",
    journal = "PASP",
    volume  = "118",
    number  = "",
    pages   = "517"
}

@article{Buitrago08,
    author  = "Buitrago F. and  Trujillo I. and Conselice C.~J. and Bouwens R.~J. and Dickinson M. and Yan H.",
    title   = "",
    year    = "2008",
    journal = "ApJ",
    volume  = "687",
    number  = "",
    pages   = "L61"
}

@article{Carnall23,
    author  = "Carnall A.~C. et~al.",
    title   = "",
    year    = "2023",
    journal = "Nature",
    volume  = "619",
    number  = "",
    pages   = "716"
}

@article{ChiosiCarraro02,
    author  = "Chiosi C. and Carraro G.",
    title   = "",
    year    = "2002",
    journal = "MNRAS",
    volume  = "335",
    number  = "",
    pages   = "335"
}

@article{Cook16,
    author  = "Cook B.~A. and Conroy C. and Pillepich A. and Rodriguez-Gomez V. and Hernquist L.",
    title   = "",
    year    = "2016",
    journal = "ApJ",
    volume  = "833",
    number  = "",
    pages   = "158"
}

@article{Eftekhari22,
    author  = "Eftekhari E. and La Barbera F. and Vazdekis A. and Allende Prieto  C. and Knowles A.~T.",
    title   = "",
    year    = "2022",
    journal = "MNRAS",
    volume  = "512",
    number  = "",
    pages   = "378"
}

@article{Feltzing01,
    author  = "Feltzing S. and Holmberg J. and Hurley J.~R.",
    title   = "",
    year    = "2001",
    journal = "A\&A",
    volume  = "377",
    number  = "",
    pages   = "911"
}

@article{FerreMateu17,
    author  = "Ferr\'e-Mateu A. and Trujillo I. and Mart{\'{\i}}n-Navarro I. and Vazdekis A. and Mezcua M. and Balcells M. and Dom{\'{\i}}nguez L.",
    title   = "",
    year    = "2017",
    journal = "MNRAS",
    volume  = "467",
    number  = "",
    pages   = "1929"
}

@ARTICLE{Ferreras25,
       author = {Ferreras I. and Trevisan M. and Lahav O. and de Carvalho R.~R. and Silk, J.},
        title = "",
      journal = "MNRAS",
         year = "2025",
       volume = "540",
       number = "",
        pages = "1069",
}

@article{Glazebrook24,
    author  = "Glazebrook K. et~al.",
    title   = "",
    year    = "2024",
    journal = "Nature",
    volume  = "628",
    number  = "",
    pages   = "277"
}

@article{Greene15,
    author  = "Greene J.~E. and Janish R. and Ma C-P and McConnell N.~J. and Blakeslee J.~P. and Thomas J. and Murphy J.~D.",
    title   = "",
    year    = "2015",
    journal = "ApJ",
    volume  = "807",
    number  = "",
    pages   = "11"
}

@article{Hilz12,
    author  = "Hilz M. and Naab T. and Ostriker J.~P. and Thomas J. and Burkert A. and Jesseit R.",
    title   = "",
    year    = "2012",
    journal = "MNRAS",
    volume  = "425",
    number  = "",
    pages   = "3119"
}

@article{Johansson12,
    author  = "Johansson J. and Thomas D. and Maraston C.",
    title   = "",
    year    = "2012",
    journal = "MNRAS",
    volume  = "421",
    number  = "",
    pages   = "1908"
}

@article{Kaviraj07,
    author  = "Kaviraj S. et~al.",
    title   = "",
    year    = "2007",
    journal = "ApJS",
    volume  = "173",
    number  = "",
    pages   = "619"
}

@article{KawataGibson03,
    author  = "Kawata D. and Gibson B.~K.",
    title   = "",
    year    = "2003",
    journal = "MNRAS",
    volume  = "340",
    number  = "",
    pages   = "908"
}

@article{LaBarbera19,
    author  = "La Barbera F. et~al.",
    title   = "",
    year    = "2019",
    journal = "MNRAS",
    volume  = "489",
    number  = "",
    pages   = "4090"
}

@article{LaBarbera13,
    author  = "La Barbera F. et~al.",
    title   = "",
    year    = "2013",
    journal = "MNRAS",
    volume  = "433",
    number  = "",
    pages   = "3017"
}

@article{Lonoce15,
    author  = "Lonoce I. et~al.",
    title   = "",
    year    = "2015",
    journal = "MNRAS",
    volume  = "454",
    number  = "",
    pages   = "3912"
}

@article{MartinNavarro18,
    author  = "Mart{\'{\i}}n-Navarro I. and Vazdekis A. and Falc\'on-Barroso J. and La Barbera F. and  Yildirim A. and van de Ven G.",
    title   = "",
    year    = "2018",
    journal = "MNRAS",
    volume  = "475",
    number  = "",
    pages   = "3700"
}

@article{MartinNavarro16,
    author  = "Mart{\'{\i}}n-Navarro I. ",
    title   = "",
    year    = "2016",
    journal = "MNRAS",
    volume  = "456",
    number  = "",
    pages   = "L104"
}

@article{MartinNavarro15,
    author  = "Mart{\'{\i}}n-Navarro I. and La Barbera F. and Vazdekis A. and Ferr\'e-Mateu A. and Trujillo I. and Beasley M.~A.",
    title   = "",
    year    = "2015",
    journal = "MNRAS",
    volume  = "451",
    number  = "",
    pages   = "1081"
}

@article{Naab09,
    author  = "Naab T. and Johansson P.~H. and Ostriker J.~P.",
    title   = "",
    year    = "2009",
    journal = "ApJ",
    volume  = "699",
    number  = "",
    pages   = "L178"
}

@article{Nanni24,
    author  = "Nanni L. et~al.",
    title   = "",
    year    = "2024",
    journal = "MNRAS",
    volume  = "527",
    number  = "",
    pages   = "6419"
}

@article{Neumann21,
    author  = "Neumann J. et~al.",
    title   = "",
    year    = "2021",
    journal = "MNRAS",
    volume  = "508",
    number  = "",
    pages   = "4844"
}

@article{Oser12,
    author  = "Oser L. and Naab T. and Ostriker J.~P. and Johansson P.~H.",
    title   = "",
    year    = "2012",
    journal = "ApJ",
    volume  = "744",
    number  = "",
    pages   = "63"
}

@article{Pipino06,
    author  = "Pipino A. and Matteucci F. and Chiappini C.",
    title   = "",
    year    = "2006",
    journal = "ApJ",
    volume  = "638",
    number  = "",
    pages   = "739"
}

@article{PipinoMatteucci04,
    author  = "Pipino A. and Matteucci F.",
    title   = "",
    year    = "2004",
    journal = "MNRAS",
    volume  = "347",
    number  = "",
    pages   = "968"
}

@article{QuilisTrujillo13,
    author  = "Quilis V. and Trujillo I.",
    title   = "",
    year    = "2013",
    journal = "ApJ",
    volume  = "773",
    number  = "",
    pages   = "L8"
}

@article{Rock16,
    author  = "R{\"o}ck B. and Vazdekis A. and Ricciardelli E. and Peletier R.~F. and Knapen J.~H. and Falc\'on-Barroso J.",
    title   = "",
    year    = "2016",
    journal = "A\&A",
    volume  = "589",
    number  = "",
    pages   = "73"
}

@article{RodriguezGomez16,
    author  = "Rodriguez-Gomez V. et~al.",
    title   = "",
    year    = "2016",
    journal = "MNRAS",
    volume  = "458",
    number  = "",
    pages   = "2371"
}

@article{SalvadorRusinol21,
    author  = " Salvador-Rusi{\~n}ol N. and Beasley M.~A. and Vazdekis A. and La Barbera F.",
    title   = "",
    year    = "2021",
    journal = "MNRAS",
    volume  = "500",
    number  = "",
    pages   = "3368"
}

@article{SanchezBlazquez07,
    author  = "S\'anchez-Bl\'azquez P. and Forbes D. and Strader J. and Brodie J. and Proctor R.",
    title   = "",
    year    = "2007",
    journal = "MNRAS",
    volume  = "377",
    number  = "",
    pages   = "759"
}

@article{Silva08,
    author  = "Silva D.~R. and Kuntschner H. and Lyubenova M.",
    title   = "",
    year    = "2008",
    journal = "ApJ",
    volume  = "674",
    number  = "",
    pages   = "194"
}

@article{Stringer15,
    author  = "Stringer M. and Trujillo I. and Dalla Vecchia C. and Martinez-Valpuesta I",
    title   = "",
    year    = "2015",
    journal = "MNRAS",
    volume  = "449",
    number  = "",
    pages   = "2396"
}

@article{Trager00,
    author  = "Trager S.~C. and Faber S.~M. and Worthey G. and Gonz\'alez J.~J",
    title   = "",
    year    = "2000",
    journal = "AJ",
    volume  = "120",
    number  = "",
    pages   = "165"
}

@article{Vazdekis16,
    author  = "Vazdekis A. and Koleva~M. and Ricciardelli E. and R{\"o}ck B., Falc{\'o}n-Barroso J.",
    title   = "",
    year    = "2016",
    journal = "MNRAS",
    volume  = "463",
    number  = "",
    pages   = "340"
}

@article{Vazdekis96,
    author  = "Vazdekis A. and Casuso E. and Peletier R.~F. and Beckman J.~E.",
    title   = "",
    year    = "1996",
    journal = "ApJS",
    volume  = "106",
    number  = "",
    pages   = "307"
}

@article{Yi05,
    author  = "Yi S.~K. et~al.",
    title   = "",
    year    = "2005",
    journal = "ApJ",
    volume  = "619",
    number  = "",
    pages   = "L111"
}
\end{multicols}
\end{document}